\documentstyle[12pt,aasms4,epsf]{article}
\begin{document}

\input psfig.sty

\newcommand{\ce}[1]{\centerline{\bf{#1}} \vskip .2in}
\newcommand{\cen}[1]{\centerline{#1}}
\newcommand{\bp}{\newpage}

\def\bo{  } 
 
\newcommand{\su}{\hspace*{.1in}}
\newcommand{\sdue}{\hspace*{.2in}}
\newcommand{\stre}{\hspace*{.3in}}
\newcommand{\squa}{\hspace*{.4in}}
\newcommand{\scin}{\hspace*{.5in}}
\newcommand{\ssei}{\hspace*{.6in}}
\newcommand{\para}{\par}
\newcommand{\be}{\begin{equation}}
\newcommand{\en}{\end{equation}}
\newcommand{\sot}{\hspace*{.8in}}

\newcommand{\lop}{\stackrel{<}{\sim}}
\newcommand{\gop}{\stackrel{>}{\sim}}
 
\def\gte{\lower 0.5ex\hbox{${}\buildrel>\over\sim{}$}}
\def\lte{\lower 0.5ex\hbox{${}\buildrel<\over\sim{}$}}
 
\def\loe{\lower 0.6ex\hbox{${}\stackrel{<}{\sim}{}$}}
\def\goe{\lower 0.6ex\hbox{${}\stackrel{>}{\sim}{}$}}

\newcommand{\ggg}{$\gamma$}
\newcommand{\eee}{$e^{\pm}$}
\newcommand{\lap}{$L_{38}^{-1/3}$}
\newcommand{\ergs}{\rm \su  erg \su s^{-1}}

\newcommand{\etal}{ {\it et al.}}

%------------------------------------------------------------------
% %     REFERENCES
%
\def\jref#1 #2 #3 #4 {{\par\noindent \hangindent=3em \hangafter=1
      \advance \rightskip by 0em #1, {\it#2}, {\bf#3}, #4.\par}}
\def\rref#1{{\par\noindent \hangindent=3em \hangafter=1
      \advance \rightskip by 0em #1.\par}}
%-------------------------------------------------------------------

\newcommand{\nn}{\noindent}

\newcommand{\ci}[1]{(\cite{#1})\/}
\newcommand{\cit}[1]{\cite{#1}}
%\newcommand{\bibitem}[1]{\bibitem{#1}}
% \newcommand{\bb}[1]{\bibitem{#1}}
 
%  Style file containing Marco Tavani modifications
%
 
%redefine the Bibliography command
\renewcommand{\thebibliography}[1]{
%\section{References\@mkboth
  %{REFERENCES}{REFERENCES}}
  \list
  {[\arabic{enumi}]}{\settowidth\labelwidth{[#1]}\leftmargin\labelwidth     \advance\lef tmargin\labelsep 
   \usecounter{enumi}}
    \def\newblock{\hskip .11em plus .33em minus .07em}
    \parsep -2pt
    \itemsep \parsep
    \sloppy\clubpenalty4000\widowpenalty4000
    \sfcode`\.=1000\relax}

\newcommand{\syn}{synchrotron }

% \centerline{\Large \bf 
%  Spectral effects on t
% The brightness distribution}
\centerline{\Large \bf
% The duration/hardness dependence of inhomogeneous}
Euclidean vs. non-Euclidean}
\centerline{\Large \bf  Gamma-Ray Bursts}
\vskip .2in
\centerline{\large M. Tavani}
\centerline{\it IFCTR-CNR, via Bassini 15, I-20133 Milano (Italy)}
\centerline{\it Columbia Astrophysics Laboratory, Columbia University, 
New York, 10027 NY (USA)}

\vskip .2in

 \centerline{\bf Abstract}
 We classify gamma-ray bursts (GRBs) according
to their observed  durations
%  double-peaked (short-long) duration distribution
and physical properties of their spectra.
We find that 
% among the four  duration/hardness GRB classes we consider,  only
long/hard bursts
(of duration $T_{90} >  2.5$~s, and 
typical photon energy $E_p \goe 0.8$~MeV corresponding to
BATSE's energy fluence hardness $H^e_{32} > 3$)  show 
% a clear inhomogeneous distribution.
the strongest deviation from the
three-dimensional  Euclidean brightness distribution.
The majority of GRBs, i.e., short bursts ($T_{90} < 2.5$~s)
and long/soft bursts (with $T_{90}>2.5$~s,
and $H^e_{32} < 3$) show little, if any, deviations from
% homogeneity.
the Euclidean distribution.
These results contradict the prediction
%  are at odds with simple 
of extragalactic GRB models  that 
 the most distant bursts  should be the  most affected by
cosmological energy redshift and time-dilation (long/soft GRBs).
The strongly 
%inhomogeneous
non-Euclidean   GRB subclass 
% strong inhomogeneity
% (possibly due to lack of burst sources at large distances)
 has very hard spectra of typical  photon
energy above $\sim 1$~MeV, i.e., outside the ideal energy range
for optimal detection by BATSE. 
Selection effects due to BATSE inefficient triggering and/or
intrinsic burst properties  may  cause the strong
deviation from the Euclidean distribution for the hardest class of
long GRBs.
% The only extragalactic scenario of GRB formation
 % consistent with o
Our results imply strong luminosity and spectral
evolution of GRB sources placed at extragalactic distances.
Whether there co-exist 
 multiple GRB populations with different 
(Galactic or extragalactic) origins remains an open question.

\vskip .2in
\centerline{Submitted to the {\it Astrophysical Journal Letters}:
November 24, 1997}
\centerline{Revised: 27 January 1998; accepted: 13 February 1998}

\centerline{Subject Headings: Gamma-rays: bursts, theory}

 \newpage
 
\baselineskip 23pt

\noindent
\section{Introduction}

The brightness  distribution  of 
gamma-ray bursts (GRBs)  depends on both the
intrinsic properties of the emission and the spatial distribution
of sources.
When analyzed globally, GRBs detected by BATSE
 show a clear deviation from the
three-dimensional Euclidean brightness distribution 
 (Meegan et al. 1992, 1995).
However, recent analyses of special GRB subclasses showed
that the deviation from the Euclidean distribution
does not apply to all bursts. On the contrary, `soft' bursts
with typical long durations appear to have an Euclidean brightness
% logN-logP
distribution (Pizzichini 1995,
Kouveliotou 1996, Belli 1997, Pendleton et al.
1997).
%  When GRB spectral sub-classes are considered,
% they appear to have different spatial distributions.
A commonly used spectral parameter is the {\it burst-averaged}  hardness ratio $H_{32}$, i.e., the ratio
of overall fluences in the BATSE
energy channels no. 2 and 3 corresponding to the
100-300~keV  and 50-100~keV bands, respectively.
The photon fluence-averaged
ratio $H^p_{32}$, and the energy fluence-averaged
ratio $H^e_{32}$ are commonly used.
A first approach was to consider
different GRB subclasses  ranked according to their $H^e_{32}$
(e.g., Pizzichini 1995).
% horizontal subdivisions of  the hardness/duration
% $H^p_{32}-T_{90}$ diagram,
% with $T_{90}$ the duration of 90\% energy deposition in the
% 50-300~keV band. 
 The analysis of Kouveliotou
et~al. (1996) found that GRBs with
photon-fluence-averaged  $H^p_{32} > 4$ and
$H^p_{32} < 2$ are distributed quite differently, with the 
soft bursts showing an Euclidean
% apparently homogeneous 
distribution.
Belli (1997) used a different phenomenological criterion
to distinguish  GRB classes in the hardness/duration
diagram ($H^e_{32}$ vs. $T_{90}$, 
with $T_{90}$ the duration of 90\% energy deposition in the
 50-300~keV band)
divided by the relation $H^e_{32} = 2 \, T_{90}^{1/2}$.
The long-duration subclass with $H^e_{32} > 4.5$ shows the 
most pronounced
  % inhomogeneous brightness distribution.
deviation from the Euclidean distribution.

Pendleton et al. (1997, hereafter P97) used a very different approach
reaching qualitatively similar conclusions.
 P97 select  GRB {\it individual pulses} 
according to the existence or lack of
a high energy component above 300~keV (BATSE's channel no. 4).
Two classes of GRBs  with without  a high energy component (HE
and NHE) were identified. Very often, 
complex GRBs with multiple pulses show both kinds of HE and NHE pulses. The interesting conclusion
of this analysis is that 
NHE pulses, both individual or as part of more complex
GRBs, appear to 
have Euclidean distributions (P97). 
%be homogeneously distributed (P97).

All of these analyses were carried out on a purely
phenomenological level, with no direct correspondence
between the selecting criterion and a physical model of emission.
The main motivation of our work is to 
provide a physical model for the
GRB observable quantities ($T_{90}, H_{32}$)
to better understand the GRB brightness distribution.
%mt3 BATSE's selection effects and
% to carry out the analysis of the brightness distribution
% with a physical interpretation. 
% For our interpretation, w
We use the shock synchrotron  model (SSM) of GRB
emission that successfully explains the great
majority of GRB burst-averaged spectra (Tavani 1996a,b, 1997;
hereafter T96a,b, T97).
The results of our GRB analysis are obviously independent of the SSM:
only the interpretation of the adopted GRB selection 
depends  on the theoretical model.
We studied GRBs of the Fourth BATSE (4B) Catalog
(Paciesas et al. 1997).
%%%%%
%
Three effects should be considered:
(1) the  intrinsic spectral
evolution within bursts (typically hard-to-soft)
that tends to concentrate bursts diagonally from top left
to bottom right of the $H^e_{32}-T_{90}$ diagram;
(2) the  effect  due to cosmological redshift, or any other
cooling mechanism simultaneously dilating the time profile and
softening the spectrum (T96a,b);
(3) selection effects due to BATSE sensitivity.
If the majority of GRB sources are at extragalactic distances,
% and well approximated by standard candles,
an overall softnening of dim and time-dilated bursts is
expected as these bursts should be the most distant for no
spectral cosmological evolution. 
The population of soft and long duration  GRBs 
is
%mt3  then
 expected 
% in simplest extragalactic models
 % to be produced at larger distances and therefore
 %mt3 to show an
be  inhomogeneously  distributed  if BATSE's sampling 
radius  exceeds the spatial distance of peak burst formation rate. 
As we show below, this is the opposite of what we find.

\section{Physical interpretation of $T_{90}$ and $H_{32}$}

The double-peaked  distribution of GRB durations is
well known (Fishman \& Meegan 1995), and
probably reflects different manifestations of 
simple (short) and complex (long) burst emission.
Pulse analysis of complex bursts shows a
%mt3 n underlying
structure of pulses with average durations 
of
%mt3  several hundred milliseconds
$\sim 200-400$~ms
 (Norris et al. 1996).
The subclass of short bursts with $T_{90} < 2.5$~s
can be modelled  as relatively simple bursts whose substructure
%mt3 
 consists
% ent with that
 of a single (or a few) pulse(s)  of
 average properties.
%  (duration and morphology).
Longer bursts ($T_{90} > 2.5$~s) are either apparently 
dilated versions of simple short bursts, or
complex superpositions of many
pulses of average properties. 
The double-peaked  $T_{90}$ distribution can 
%mt3 therefore
then
be related to the relative complexity of the pulse structure.
No apparent correlation  exists between overall burst durations and peak photon intensities. We  define
here  `short' and `long'  bursts
divided by the critical duration $T_{90} = 2.5$~s 
corresponding to the local minimum of the $T_{90}$ distribution.

There is instead a definite correlation between $T_{90}$ and
average GRB hardness, most likely  as a consequence of the
hard-to-soft spectral  evolution within bursts (e.g.,
Ford et al., 1995). Fig.~1 shows the 4B
hardness/duration diagram ($H^e_{32}$ vs. $T_{90}$)
with a clear indication that short bursts are generally
harder than long ones. Spectral softnening for individual
pulses and for long
bursts has been reported in many cases (Ford et al 1995), and it
constitutes a fundamental property of cooling
particle distributions (e.g., Tavani, 1998).

We apply the synchrotron shock model (T96a,b)
for the interpretation of the hardness ratio $H^e_{32}$.
The SSM is based on a hydrodynamic and radiative model of
relativistic MHD outflows interacting with shocks that
reconvert part of their energy in impulsive acceleration.
The SSM can in principle
 be applied to relativistic mass/electromagnetic outflow models
at extragalactic and Galactic distances.
Average  hardness ratios $H^e_{32}$
 can be reliably converted into  { average
 peak energies, $E_p$, of the $\nu \, F_{\nu} $ spectrum } (T97).
Fig.~3 of T97 shows the SSM calculated $H^e_{32}$ 
(and $H^p_{32}$) as a function of $E_p$
for different assumptions about the index 
$\delta$ of the high-energy component of the post-acceleration
particle energy distribution function ($N(\gamma) \propto 
\gamma^{-\delta}$). The SSM calculated $H^e_{32}$'s are in
good agreement with detailed broad-band average
spectra of GRBs (T96a,b;T97). The agreement between theory and
observations  extends also to GRBs detected above 1~MeV (T96a,b).
%
 % rimosso
%
% As an example of a burst detected
% from 30~keV to hundreds of MeV, we can consider
 % the  bright GRB~910503 with $E_p \simeq
% 0.8$~MeV determined by a CGRO multi-instrument (not only BATSE)
% fit  (Schaefer et al. 1996, T96a).
% BATSE's tabulated average $H^e_{32} = 6.01 \pm 0.04$ (Meegan et al., 1995) % agrees  within
a factor of $\sim 2$ with the SSM calculated value
% from a CGRO multi-instrument fit\footnote{
% The lack of perfect agreement between  SSM's and  BATSE's
% hardness ratio $H^e_{32}$ in the case of GRB~910503 
% despite the  overall excellent SSM  fit of the CGRO  multi-instrument % spectrum is clearly  due to
an apparent depletion of soft photons
%%  affecting the determination of  $H^e_{32}$.}
% for $E_p = 0.8$~MeV (T96a).
% 
% 
 %%   143  57.236e-07  34.374e-09  75.901e-07  43.080e-09
% %        45.620e-06  13.972e-08  23.324e-05  12.379e-07
 %%           52.008       1.426       1.280
 %%           47.569       0.682       1.344
 %%           37.634       0.301       2.880
 %
 %
 We can therefore physically interpret the average hardness
 ratio by converting its value to a typical photon energy $E_p$ 
 at which most of the burst energy is dissipated.
 We choose $H^e_{32} = 3$ as the critical value
 distinguishing GRBs with typical photon energies
 below or above 0.8~MeV. We call them `soft' and `hard'
 bursts, respectively.
 We also carried out calculations for different choices
 of the critical $H^e_{32,c}$, in the range
 $2.5 \loe H^e_{32,c} \loe 4$ that would correspond to different
 peak energies ($\sim 400$~keV and $\sim 2$~MeV, respectively;
 see Fig.~3 of T97).
Our main conclusions do not depend on the choice of
% the 
%mt3 critical $H^e_{32}$ 
$H^e_{32,c}$
in the range given above.
Our preferred choice ($HR^e_{32,c} = 3$) is physically
justified by selecting GRBs whose typical $E_p$ is
%mt3 approximately four
$\sim  4$~times  the average $E_p$ detected by
BATSE ($\sim 200$~keV, Band et al. 1993), and 
outside the optimal energy range for BATSE detection
(50-300~keV). The energy spectrum of GRB~910503 
Schaefer et al. 1996,T96a)  clearly shows the spectral properties
of GRBs corresponding to our critical value for $H^e_{32,c}$.
The range for ideal BATSE detection is  $1 \loe H^e_{32} \loe 2.5$.

Fig.~1 clearly shows that short GRBs have average $E_p$ well
above 1~MeV, and are efficiently detected by a combination of
short risetimes (below 1~s) and spectral evolution in
the BATSE's optimal detection range.
We notice the relative small number of short/soft GRBs,
contrary to the large number of long/soft bursts.
This feature is  not influenced by selection
effects, and  reflects  the GRB hard-to-soft
spectral evolution from simple GRBs to multiple-pulsed events.
Selection effects may play a role in determining the
trigger efficiency for long/hard events populating the
upper right corner of the hardness/duration diagram.
In our analysis, we consider only bursts with well
defined spectral properties, including only events with
relative uncertainty on $H^e_{32}$ less than 50\%.
We also note that only $\sim 3$\% of the events in Fig.~1
show a deviation of their measured $H^e_{32}$ by more than $3\sigma$
larger than the ideal maximum allowed SSM value ($H^e_{32}\simeq 5$, see T97). These events are
most likely characterized by
soft photon depletion/absorption in the 50-300~keV band
that may  influence the observed $H^e_{32}$ in addition to
instrumental effects.

\section{Analysis of the 4B catalog}

Equipped with a physical interpretation of the $H^e_{32}-T_{90}$ diagram, we analyzed the brightness
distribution of 4B GRBs according
to their different hardness ratios and durations.
We distinguish four classes:\\
(S/H) GRBs with $T_{90} < 2.5$ sec, and $H^e_{32} > 3$ (short/hard bursts);\\ (S/S) GRBs with
$T_{90} < 2.5$ sec, and $H^e_{32} < 3$ (short/soft bursts);\\ (L/H) GRBs with $T_{90} > 2.5$ sec,
and $H^e_{32} > 3$ (long/hard bursts);\\ (L/S) GRBs with $T_{90} > 2.5$ sec, and $H^e_{32} < 3$
(long/soft bursts).

Fig.~2 shows the results, where we plot the cumulative burst 
number with peak photon intensity above
a given value $P_{64}$ as a function of
$P_{64}$ (64~ms trigger timescale).
Both classes L/S and S/H show brightness distributions 
% consistent with being homogeneous and 
in {\bo good}
 agreement with the  $-3/2$ Euclidean slope.
The deviations from the Euclidean distribution are minor,
with a discrepancy below a factor of 2
 between real and inferred values from the high-$P_{64}$'s
at $P_{64} = 1.3 \rm \, ph. \, cm^{-2}  \, s^{-1}$.
The curvature  near  
$P_{64} \loe 1 \rm  \,ph. \, cm^{-2}  \, s^{-1}$ is clearly due to
a selection effect.
{\bo 
Both L/S and S/H distributions
might also be marginally
consistent with a  high-flux slope of --3/2 followed by 
an intermediate-flux slope near --1 for $P_{64} \loe 8
\rm \, ph. \, cm^{-2}  \, s^{-1}$. }
% ($P_{64}$ is the peak photon intensity for the 64~ms 
% trigger timescale).
 On the contrary, the
 long/hard (L/H) bursts
 show a strong deviation from the
Euclidean  distribution
at  $P_{64} \simeq  20 \rm \, ph. \, cm^{-2} \,
s^{-1}$.  The fourth class of short/soft events contains very few
(66) and faint events and its brightness distribution is 
not relevant here\footnote{ If we summed
together all the soft and hard events with
 $T_{90} < 2.5$~s we would obtain results very similar to those for
the S/H class.}.
These results are confirmed  also for brightness
distributions obtained for  256 and 1024~ms trigger timescales.

All four duration/hardness GRB   classes show 
% highly 
isotropic sky distributions.
Table~1 gives the values obtained for the dipole and quadrupole
moments relative to the Galactic center and plane.
We also obtained the
 values of the $V/V_{max}$ parameter for 
GRBs with tabulated ratios $C_p/C_{min}$ (with
$C_p$ and $C_{min}$ the peak and threshold  countrates, respectively
for the 64~ms trigger timescale in the BATSE's energy range).
%mt3 Fig.~3 shows the $V/V_{max}$ distributions for the three
% relevant duration/hardness GRB classes, and 
Table~1 reports the average values of $V/V_{max}$
for the three duration/hardness GRB classes.
%mt2
{\bo  We note that a spectral bias in the determination of
$C_{min}$ is to be expected (see below), and that tabulated values of
$C_{min}$ can be lower limits to the actual minimum countrates
for detection of bursts with 
 duration/spectral properties  outside BATSE's optimal range. }
 %can be detected by BATSE. }

Our results  indicate that only
%mt3  the subclass of
long/hard  GRBs shows a strong
%  inhomogeneous
deviation from the Euclidean brigthness  distribution. 
These events  dominate the bright end of the GRB population 
(one out of two GRBs are L/H bursts  in the range 
$6 {\rm \, ph. \, cm^{-2}  \, s^{-1}} \loe
P_{64} \loe  100 \rm  \, ph. \, cm^{-2}  \, s^{-1}$) and yet
they show the strongest deviation from the Euclidean slope.

\section{Discussion}

We briefly discuss here some points relevant to the interpretation
of our results, postponing a more thorough discussion to
a later publication.
Our results confirm previous analyses of the GRB brightness
distribution, and contradict  the predictions 
of simple cosmological models 
with no strong spectral and luminosity evolution.
The   total sample of
GRBs detected by BATSE
shows a clear deviation  at low peak intensities
% (near $10 \rm ph \, cm^{-2} \, s^{-1}$ at 64~ms integration
% time) 
from the Euclidean brightness distribution. This property,
together with an average $V/V_{max}$ value near 0.37
(see Table~1)
 was interpreted as a signature of
 a globally inhomogeneous distribution in space
(e.g., Paczynski 1991, Briggs 1996).
%mt2  cambio qui 
  However, we find that 
% these conclusions  apply
this conclusion can be applied 
 only to the special subclass of long/hard bursts. 
 The fact that
%  that long and hard
{\it only 
%mt2 L/H 
long/hard GRBs 
% deviate from the Euclidean brightness distribution
% and that they are on the average
are strongly non-Euclidean being at the same time
 very intense}  is puzzling for  extragalactic models.
%mt2
{\bo  The soft and long GRBs that would naturally exhibit
an inhomogeneous brightness distribution in simple cosmological
models  show little, if any, deviations from being Euclidean.
We emphasize that our conclusion is deduced by the
{\it combination} of average spectral and 
% duration
time dilation
 properties of GRBs.
Furthermore,}
the peak intensity distributions of  S/H, L/S and L/H bursts
are similar to each other, 
%mt2 excluding
{\bo making unnatural }
 that  L/H bursts are in
the foreground of more distant (and possibly softer) bursts.
We are left with the conclusion that strong luminosity and
spectral cosmological evolution 
%mt2 of GRBs 
is necessary to
reconcile our results with
extragalactic models of GRBs.
%mt3  formation.
%mt3   models  placing the majority of 
%mt2 bursts
% GRBs   at extragalactic distances.
 L/H bursts placed in remote regions where 
a relative decrease of their spatial density 
% causes the strong deviation
makes them strongly non-Euclidean 
% from the Euclidean brightness distribution
 would have to be on the average very bright
(by a factor of several) and  with harder spectra
($E_p$  larger by a factor of $\sim 10$) than the
majority of closer bursts.
%  which are presumably closer, and therefore
% substantially less luminous and softer. 
Whether this can be
accomplished by  GRB formation  mechanisms in
star forming galaxies (SFGs) (as suggested by recent
HST observations of GRB~970508, Pian et al. 1998)
remains to be seen.
The redshift dependence of  SFG density  (Madau et al. 1996)
is strongly peaked near $z \simeq 1$ and then decreases
for larger redshifts. L/H bursts might originate at
redshifts $z \goe 1$ under the condition that more distant GRB 
sources produce bursts more luminous and harder than those at
smaller redshifts.
The $V/V_{max}$ distribution of 
L/H bursts interpreted at face value with no
selection bias,  
%mt3 (see Fig.~3),
 indicate that these
bursts  tend to reside in relatively close regions. 
The May 8th 1997 burst detected by BSAX (Piro et al. 1997)
has an equivalent hardness $H^e_{32} = 3.6 \pm 0.9$ and
$T_{90} = 8$~s
(Amati et al. 1997).
% , personal communication).
If GRB~970508  belongs to the L/H subclass of GRBs, 
 %mt2 it 
is in the 
strongly non-Euclidean part of the brightness distribution.
%mt2 and o
Our results imply that, 
{\bo if the optical transient at cosmological distance is
associated with the burst, GRB~970508 might be}
%  it should be 
located at 
a redshift larger than $z=0.83$ (indicated by optical spectroscopy, 
%mt2 of the associated optical transient, see 
Metzger et al. 1997).

Alternative interpretations  of our results should also be considered.
One possibility 
that reconciles our results with those of 
%mt2 P97
{\bo Pendleton et~al. (1997)}
is that BATSE selection effects play a major
role in shaping the brightness distribution of duration/hardness
subclasses of GRBs.
The ability by BATSE to trigger on L/H events
 might crucially
depends on their  peak intensity (see also 
Lloyd \& Petrosian 1997).
 As clearly indicated by the
P97 results, even single pulses within complex GRBs share
the same characteristics of our burst-averaged 
S/H, L/S and L/H subclasses. This 
%mt2 shows 
{\bo fact demonstrates }  that short 
and long/soft bursts (pulses) can be detected with good efficiency
given BATSE's  trigger conditions.
However, detection of relatively
long and hard bursts (or L/H pulses usually occurring at
the beginning of complex GRBs, see P97) require special
conditions. It is conceivable that these
trigger conditions are favorably met for quite bright GRBs
above $P_{64} \goe  10 \rm \,  ph. \, cm^{-2}  \, s^{-1}$.
Detecting a burst (pulse) which is  long/hard 
(or with substantial flux above 300 keV) may
be difficult for smaller values of $P_{64}$.
This might explain the sharp deviation from the Euclidean
slope 
%mt2 
of the L/H burst brightness distribution.
Since BATSE detects short/hard bursts with no apparent trigger
inefficiency for low peak intensities (see Fig.~2),
we conclude that long duration bursts (possibly with slow risetimes)
with typical $E_p$'s  outside {\bo BATSE's optimal detection range}
%mt2 the ideal range for  detection by BATSE
are  the most affected by the trigger criteria.
The results of Lloyd \& Petrosian (1997) 
{\bo indicating}
 a corrected
$E_p$ distribution with a large undetected population of GRBs 
% (by a factor $\goe 3$)
 with $E_p \goe 1$~MeV  agree with this interpretation.
{\bo BATSE's trigger conditions [flux enhancement over
background nominally by 5.5~standard deviations
% $\sigma$ 
within a given
timescale (64, 256 and 1024~ms) and for the energy range 50-300~keV]
strongly favors detection of short bursts as well as long bursts with
typical emission in the optimal energy range (our L/S bursts).
Slow risetime-bursts with average $E_p \goe 1$~MeV and no strong
hard-to-soft spectral evolution are 
%mt3 clearly
 disfavored for BATSE's detection. 
{\it Both} burst-averaged and individual-pulse
properties of GRB brightness distributions share the same
Euclidean (non-Euclidean) character of their soft (long/hard) 
% burst (pulse)
population.
This fact finds a natural interpretation
in  BATSE's selection effects. 
%mt3 influencing the logN-logP distribution.
%mt2 may be  affected by   BATSE's selection effects.
} 
%mt2  Our results based on burst-averaged properties
% and those  obtained by P97 for individual GRB pulses 
% can therefore be interpreted in a consistent way.

%mt2
{\bo 
We also note that BATSE  spectral selection effects 
can strongly influence the actual  values of $C_{min}$.
 % tabulated for the optimum range of 50-300~keV.
Taken at face values, the average $V/V_{max}$'s  obtained for
our duration/spectral GRB  classes (Table~1) indicate
 inhomogeneous distributions by more than three standard deviations
for all except the S/S bursts.
However, BATSE's  trigger conditions
%mt3  spectral response
%mt3  dependent on average burst intensity 
can couple the distance and spectral dependence of 
countrates in a non-trivial way, breaking the
$1/r^2$ dependence of $C_p$ deduced in previous 
applications of the $V/V_{max}$ method to GRBs
(e.g., Schmidt, Higdon \& Hueter, 1988).
The nominal values  of $C_{min}$ tabulated
 in the 50-300~keV range can underestimate
(by a factor of a few)  the actual values  of $C_{min}$
necessary to detect bursts well outside BATSE's  optimal
spectral and risetime ranges.
The  long/hard and short/hard GRB populations 
are  the most affected by this bias.}
%mt2  The brigthness and $V/V_{max}$ distributions for special subclasses
 % of GRBs might  be consistent with homogeneity.}

% If selection effects induce the deviation from the Euclidean 
% distribution % of L/H GRBs one can
% reconsider also Galactic models.
Disregarding a Galactic origin for GRBs might have been
premature.
The combination of sky isotropy, inhomogeneity and 
the  average value of $V/V_{max}  < 1/2$
  applied to the whole sample of  GRBs
was interpreted as being inconsistent with 
standard Galactic models
(e.g., Paczynski 1991). However,
our results clearly indicate that the strong deviation from
a homogeneous distribution (if that is the correct interpretation
of non-Euclidean  GRBs) applies  only to long/hard bursts,
and certainly not to the majority of GRBs.
A Galactic  origin of a subclass or the majority
of  GRBs is not excluded by our results.

\vskip .1in
The author thanks an anonymous referee for helpful comments.

 \newpage

\vskip .3in
\noindent
{\bf References}

\noindent
Amati L., and the BSAX Team, 1997, personal communication.

\noindent
Band,  D., et al., 1993, ApJ, 413, 281.

\noindent
Belli, B.M., 1997, ApJ, 479, L31.

\noindent
Briggs, M., et al., 1996, ApJ, 459, 40. 

\noindent
Fishman, G.J. \& Meegan, C.A., 1995, ARA\&A, 33, 415.

\noindent
Ford. L.A. e al., 1995, ApJ, 439, 307.

\noindent
Kouveliotou C., et al., 1996, AIP Conf. Proc. no. 384, p. 42.

\rref{Lloyd, N.M. \& Petrosian, V., 1997,
in the Proceedings of the 4th
       Huntsville Gamma Ray Burst Symposium, 
eds. C.A. Meegan, P. Cushman (New York: AIP), in press
(astro-ph/9711192)}

\noindent
Madau, P. et al., 1996, MNRAS, 283, 1388.
 
\noindent
Meegan C.A., et al., 1992, Nature, 355, 143.

\noindent
Meegan, C.A., et al., 1995, 3B catalog, ApJS, 106, 65.

\noindent
Metzger, M.L., et al., 1997, Nature, 387, 878.

\noindent
Norris, J.P. et al., 1996, ApJ, 459, 393.

\rref{
Paciesas, W.S., et al., 1997,
The Fourth BATSE Gamma-Ray Burst Catalog,
http://www.batse.msfc.nasa.gov/data/grb/4bcatalog}

\noindent
Pendleton, G. et al., 1997, ApJ, in press (P97).

\noindent
\rref{Pian E. et al., 1998, ApJ, 492, L103}

\noindent
Piro L., et al., 1997, A\&A, submitted.

\rref{Pizzichini, G., 1995, in the Proc. of the XXIV ICRC Conference,
p. 81}

\rref{
Schaefer B. et al., 1996, in AIP Conf. Proc. Series no. 384, p. 274}

\rref{Schmidt, M., Higdon, J.C. \& Hueter, G., 1988, ApJ, 329, L85}

\noindent
Tavani, M., 1996a, Phys. Rev. Letters, 76, 3478 (T96a).

\noindent
Tavani, M., 1996b, ApJ,  466, 768 (T96b).

\noindent
Tavani, M., 1997, ApJ, 479, 135 (T97).

\rref{Tavani, M., 1998, 
in the Proceedings of the 4th Huntsville Gamma Ray Burst Symposium, 
eds. C.A. Meegan, P. Cushman (New York: AIP), in press}

\newpage

\vspace*{.7in}

% \begin{table}
\begin{center}
{\bf Table 1: Properties of duration/hardness GRB subclasses}
\vskip .05in
\begin{tabular}{|l|lll|}
\hline
GRB duration/hardness class  & $<\cos(\theta)>$ &
$<\sin^2(b) - 1/3>$  & $<V/V_{max}>$ \\
\hline
long/soft (520) & $-0.002 \pm 0.043$ & $-0.007 \pm 0.013$&
$0.434 \pm 0.019$  (221)\\
long/hard (338) & $-0.020 \pm 0.054$ & $+0.008 \pm 0.016$&
$0.289 \pm 0.022$ (168)\\
short/hard (254)& $-0.062 \pm 0.063$ & $-0.019 \pm 0.018$&
$0.354 \pm 0.023$ (149)\\
short/soft (66) & $+0.003 \pm 0.123$ & $+0.042 \pm 0.036$&
$0.458 \pm 0.044$ (43)\\
total (1178)    & $-0.019 \pm 0.016$ & $-0.007 \pm 0.013$&
$0.373 \pm 0.012$ (581)\\
\hline
\end{tabular}
\end{center}
\vskip .1in
% The table gives the dipole, quadrupole moments with
% respect to the Galactic center and plane, and the average
% value of $V/V_{max}$. We follow the definitions of 
% Briggs et al. (1996).
% The angle $\theta$ is the angle between the burst 
% position and the Galactic center.
% The angle $b$ is the Galactic latitude.
% \end{table}

\vskip .3in

% \vskip .5in
% \ce{Table Caption}

The table gives the dipole, quadrupole moments with
respect to the Galactic center and plane, and the average
value of $V/V_{max}$. We follow the definitions of
Briggs et al. (1996).
The angle $\theta$ is the angle between the burst
position and the Galactic center.
The angle $b$ is the Galactic latitude.

% \end{document}

\newpage

\begin{figure}
\epsfxsize=16cm
\epsfysize=16cm
\vspace*{-2.cm}
\centerline{\epsffile{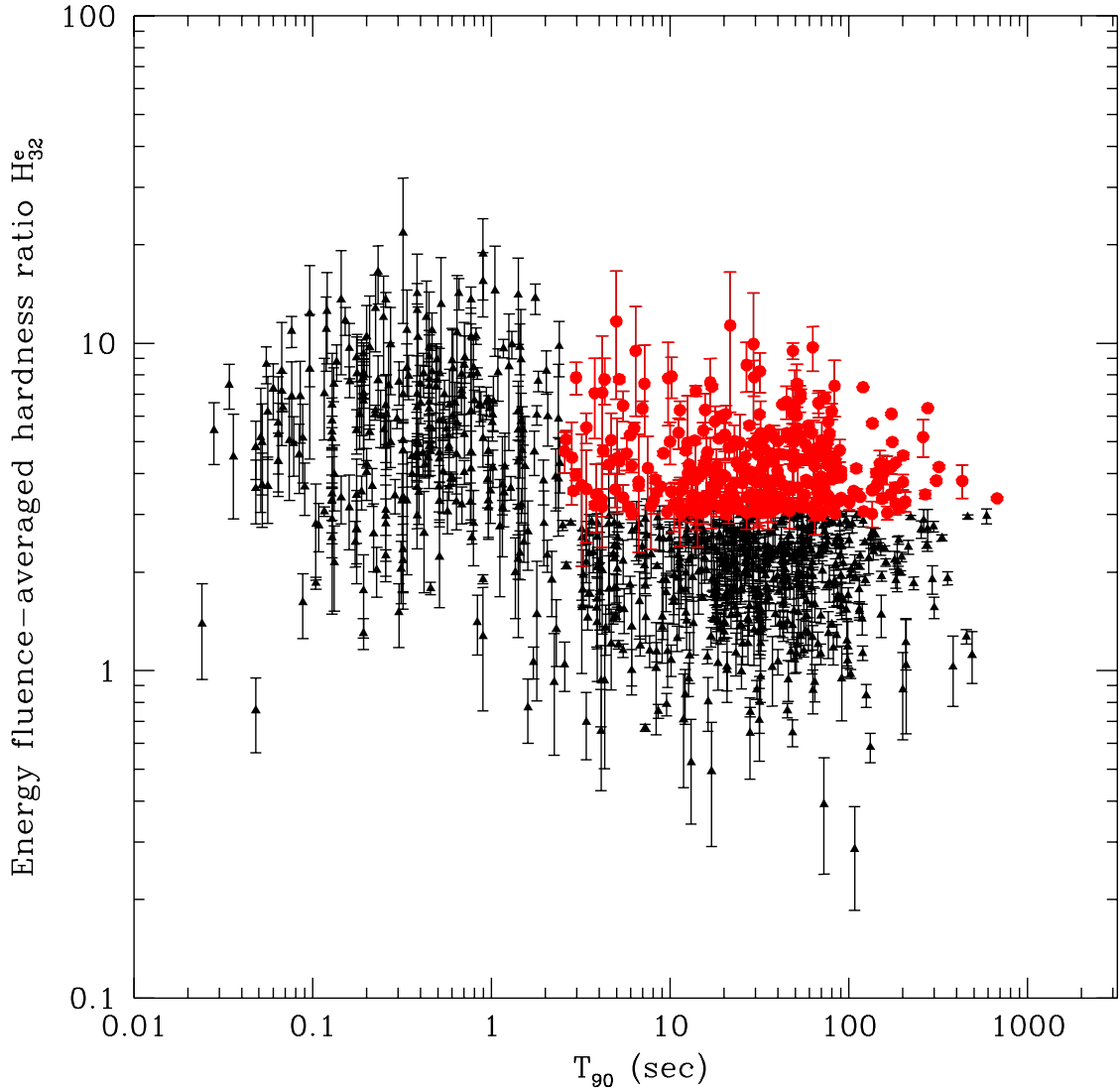}}
\vspace*{-1.cm}
\caption[image]{
GRB energy  fluence-averaged  hardness ratio ($H^e_{32}$) vs. duration
($T_{90}$) distribution from the 4th BATSE catalog (4B).
Only GRBs with fractional error on $H^e_{32}$ less than 50\% are
plotted. Error bars are $1\sigma$.
% The population of l
Long/hard (L/H) GRBs 
% characterized by a strong inhomogeneous brightness distribution
are  marked by solid filled circles:
they show the strongest deviation from the Euclidean brightness 
distribution (see Fig.~2).
% The rest of GRBs appear to have homogeneous brightness distributions
% (see Fig.~2).
% In particular, the long/soft (L/S) bursts which constitute the 
% majority of GRBs in the 4B sample are expected to be the 
% most affected by cosmological redshift and time-dilation
% effects if they originate from a distance larger than for the
% other GRBs.
}
\end{figure}

\begin{figure}
\epsfxsize=16cm
\epsfysize=16cm
\vspace*{-2.cm}
\centerline{\epsffile{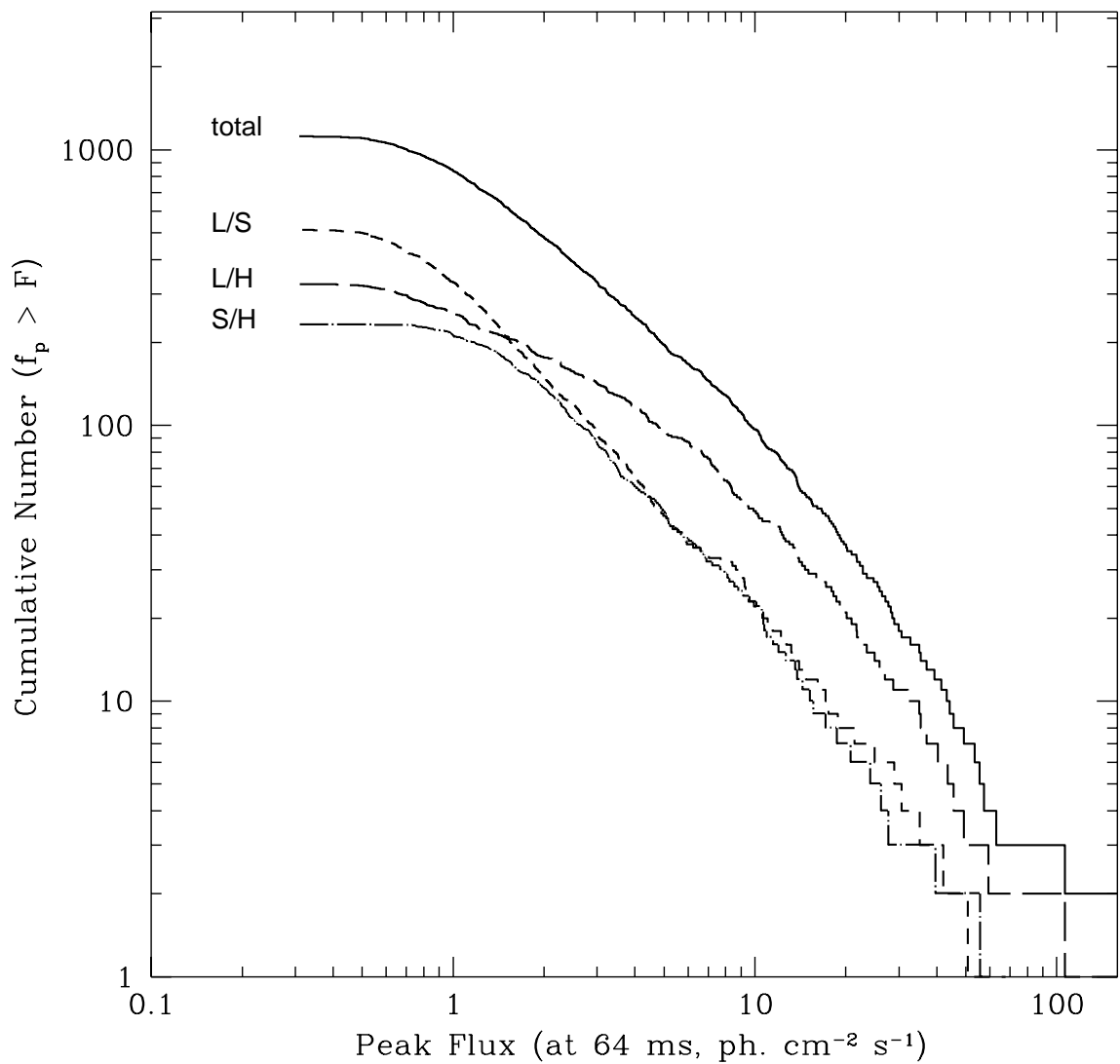}}
\vspace*{-.5cm}
\caption[image]{
Brightness distribution  (cumulative number of bursts $N$ above
a certain peak photon intensity level $P_{64} $ as a function of $P_{64}$ (64~ms trigger timescale)
for different subclasses of GRBs.
{\it Solid histogram:} total distribution;
{\it short-dashed histogram:} long/soft bursts;
{\it long-dashed histogram:} long/hard bursts;
{\it dotted-dashed histogram:} short/hard bursts.
}
\end{figure}

\end{document}